\documentclass[12pt,preprint]{aastex}

\shorttitle{Compact jets in a neutron star X-ray binary}
\shortauthors{Migliari et al.}

\begin{document}


\title{{\it Spitzer} Reveals Infrared Optically-Thin Synchrotron Emission from the Compact Jet of the Neutron Star X-Ray Binary 4U 0614+091}


\author{S. Migliari\altaffilmark{1}, J.A. Tomsick\altaffilmark{1}, T.J. Maccarone\altaffilmark{2}, E. Gallo\altaffilmark{3}, R.P. Fender\altaffilmark{2}, 
G. Nelemans\altaffilmark{4}, D.M. Russell\altaffilmark{2}}
%\affil{Center for Astrophysics and Space Sciences, University of California San  Diego, 9500 Gilman Dr., La Jolla, CA 92093-0424 }

%\author{T. J. Maccarone\altaffilmark{2}}
%\affil{National Optical Astronomy Observatories, Tucson, AZ 85719}
%\email{aastex-help@aas.org}

%\and

%\author{R. J. Hanisch\altaffilmark{5}}
%\affil{Space Telescope Science Institute, Baltimore, MD 21218}


\altaffiltext{1}{Center for Astrophysics and Space Sciences, University of California San  Diego, 9500 Gilman Dr., La Jolla, CA 92093-0424}
\altaffiltext{2}{ School of Physics and Astronomy, University of Southampton, Hampshire SO17 1BJ, United Kingdom}
\altaffiltext{3}{Chandra Fellow, Physics Department, University of California Santa Barbara, CA  93106-9530}
\altaffiltext{4}{ Department of Astrophysics, Radboud University Nijmegen, PO Box 9010 6500 GL, Nijmegen, The Netherlands}


\begin{abstract}
{\it Spitzer} observations of the neutron star (ultra-compact) X-ray binary (XRB) 4U 0614+091  with the Infrared Array Camera reveal emission of non-thermal origin in the range 3.5-8 $\mu$m. The mid-infrared spectrum is well fit by a power law with spectral index of $\alpha=-0.57\pm0.04$ (where the flux density is $F_{\nu}\propto\nu^{\alpha}$). 
Given the ultra-compact nature of the binary system, we exclude the possibility that either the companion star or the accretion disc can be the origin of the observed emission. 
These observations represent the first spectral evidence for a compact jet in a low-luminosity neutron star XRB and furthermore of the presence, already observed in two black hole (BH) XRBs, of a  `break' in the synchrotron spectrum of such compact jets. We can derive a firm upper limit on the break frequency of the spectrum of $\nu_{thin}=3.7\times10^{13}$~Hz, which is lower than that observed in BH XRBs by at least a factor of 10. 
Assuming a high-energy cooling cutoff at $\sim1$~keV, we estimate a total (integrated up to X-rays) jet power to X-ray bolometric luminosity ratio of $\sim5\%$, much lower than that inferred in BHs.
%We briefly discuss the possible implications of the differences in jet-to-X-ray bolometric power and $\nu_{thin}$ between 4U~0614+091  and the BH GX~339-4 in the advection and jet-dominated frameworks.
\end{abstract} 





\section{Introduction}

For X-ray binaries (XRBs), the radio band has always been a privileged window for studies of relativistic  jets. This is because, neither the companion star nor the accretion disc emits significant radiation at very long wavelengths and the synchrotron emission from the jet dominates. The radio power in XRBs is observed to be strictly related to the X-ray emission (see Fender 2006 for a review). In black hole (BH) XRBs, the hard state is associated with a radio active state (Fender 2001), while in the thermal dominant state the radio emission is quenched (Fender et al. 1999; see e.g. McClintock \& Remillard 2006 for a definition of the X-ray spectral states). The radio emission in hard state BHs is characterised by an optically thick synchrotron radio spectrum (i.e. $\alpha\ge0$ where $F_{\nu}\propto\nu^{\alpha}$ and $F_{\nu}$ is the flux density at a frequency $\nu$; e.g. Fender 2001). This optically thick synchrotron emission is interpreted as a spectral signature of conical continuously replenished, self-absorbed compact jets (Blandford \& K\"onigl 1979; Hjellming \& Johnston 1988; Kaiser 2006). This interpretation has been confirmed by radio imaging of the milli-arcsecond scale jets, for two BH XRBs: Cyg X-1 (Stirling et al. 2001) and GRS~1915+105 (although not in a canonical hard state; e.g. Dhawan, Mirabel \& Rodr\'iguez 1999).    
At shorter wavelengths, compact jet models predict a `break' from an optically thick to an optically thin (e.g. $\alpha\sim-0.6$) synchrotron spectrum. The optically thin spectrum represents the emission from the part of the jet that is closer to the base and not self-absorbed. In only two BH XRBs, near-infrared observations have directly shown this part of the spectrum: GX~339-4 (Corbel \& Fender 2002; see also Nowak et al. 2005) and XTE J1118+480 (Hynes et al. 2003a).  For other BHs, the optically thick synchrotron spectrum ranges from the radio band to the infrared (IR) and possibly even beyond and the emission of the disc and/or the companion star precludes a direct observation of the optically thin part (e.g. Fender 2001; Hynes et al. 2000; Corbel et al. 2001). Note that the actual frequency at which the break occurs, i.e. the `knee', has not been directly  observed yet (except possibly in GX~339-4; Corbel \& Fender 2002), but only in some cases inferred by fits with broken power laws (see Nowak et al. 2005). The detection of the optically thin part of the synchrotron spectrum is fundamental in order to determine the total power carried by the jet.

Atoll neutron star (NS) XRBs are a class of low-magnetic field NS systems accreting at relatively low rates, which show two distinct X-ray states, `island' and `banana' states,  directly comparable, respectively, to the hard and thermal dominant states of BHs (see Hasinger \& van der Klis 1989; van der Klis 2006 for a review). Atolls in their hard state show radio emission that is about a factor of 30 lower than that of BH XRBs at the same observed X-ray luminosity (Migliari et al. 2003; see Fender \& Hendry 2000; Migliari \& Fender 2006). 
However, although brightness temperature arguments indicate that the radio emission comes from out-flowing matter, given the lower signal-to-noise ratio of the radio detections of atoll sources, we do not have yet unambiguous observational constrains on the spectral index of their radio spectrum (see e.g. Fender 2006; Migliari \& Fender 2006) . 
Three low-luminosity NSs, the accreting millisecond X-ray pulsars SAX~J1808.4-3658, XTE~J0929-314 and XTE~J1814-338, show an optical/near-IR excess that can be explained by a non-thermal contribution from a jet (Wang et al. 2001; Giles et al. 2005; Krauss et al. 2005).

The source 4U~0614+091  is an ultra-compact NS XRB, with a carbon-oxygen white dwarf donor (Juett, Psaltis \& Chakrabarty 2001; Nelemans et al. 2004). It has been classified as an atoll source and has been observed to be almost persistently in its hard (island) state (e.g. van Straaten et al. 2000). This last characteristic plus its relative proximity (i.e. a distance of $<3$~kpc; Brandt et al. 1992) make 4U~0614+091 the best candidate for the possible detection and imaging of its compact jet. However, to date, only a $3\sigma$ upper limit of $<0.09$~mJy has been obtained with a Westerbork Synthesis Radio Telescope (WSRT) observation at 5 GHz (Migliari \& Fender 2006). 

In this Letter, we present the spectral evidence for a compact jet in 4U 0614+091, the first in a low-luminosity NS system. We detected the IR counterpart of 4U~0614+091 with the Infrared Array Camera (IRAC) on-board the {\it Spitzer Space Telescope}, and found evidence for optically {\it thin} synchrotron radiation emitted by the non-self-absorbed part of a compact jet. These observations are part of a {\it Spitzer} survey of ultra-compact X-ray binaries, aiming to detect the IR non-thermal (jet) emission from NS binary systems; the complete results of the survey will be reported in an upcoming paper. 

\section{Observations}

We observed 4U~0614+091  with {\it Spitzer} IRAC at 3.6, 4.5, 5.8 and 8~$\mu$m, on 2005 October 25 (UT 17:50:06-17:56:36). 
We have processed the Basic Calibrated Data using the software {\tt mopex} (Makovoz \& Marleau 2005).
%, following the procedure as in the {\it Spitzer} Science Center (SSC) manuals\footnote{Available on line at the SSC web-page http://ssc.spitzer.caltech.edu}. 
We created a mosaic from the 10 frames per band obtained in the observations, created a point-response function (PRF) with {\tt prf\_estimate}  and extracted the source flux from the background subtracted image using {\tt apex}.  We have corrected for interstellar extinction using 
$A_{\rm v}=2$ [derived from the equivalent hydrogen column density values in Juett, Psaltis \& Chakrabarty (2001) and using 
$N_{\rm H}=A_{\rm v}\times0.179\times10^{21}$  in Predehl \& Schmitt (1995)] and following the standard optical-to-IR interstellar extinction law (e.g. Rieke \& Lebofsky 1985; Cardelli, Clayton \& Mathis 1989). The corrections for interstellar extinction are small, i.e. $\sim10\%$ for the flux density at $3.6 \mu$m and less than 5\% for the flux densities in the other three IRAC bands. We added a $5\%$ systematic error on the estimate of the flux densities to take in to account the uncertainties on the photometric calibration (see also Reach et al. 2005).

\section{Results and discussion}

We detected the IR counterpart of 4U~0614+091  in all four IRAC bands.  
In Fig.~1, we show the 4.5~$\mu$m image of 4U~0614+091: a fit of the source with the PRF gives coordinates R.A.=$6h17m07s.35(3)$ and Dec.= $+09^{\circ}08^{\prime}13\arcsec.60(5)$, coincident with the position of the optical counterpart V1055 Ori ({\it cross}). In Fig.~2 we show the IRAC spectrum of 4U~0614+091 ({\it filled circles}) together with the non-simultaneous observed mean flux density of the optical counterpart ({\it star}) and near-IR {\it J, H, K} United Kingdom Infrared Telescope (UKIRT) observations ({\it triangles}; D.M. Russell et al. in preparation). 
The IRAC observations are well fit by a power law with $\alpha=-0.57\pm0.04$ ({\it solid line}), which is consistent with the optically thin synchrotron emission observed from jets in XRBs. The optical counterpart has been observed to vary by $\sim0.5$ mag with respect to the mean (Machin et al. 1990; the vertical bars reflect this variation), and its spectrum is consistent with thermal emission from the accretion disc (Machin et al. 1990; see also Nelemans et al. 2004 and Juett et al. 2001). 

In ultra-compact XRBs, the thermal spectrum of the disc in the optical and IR bands is expected to follow the Rayleigh-Jeans law $F_{\nu}\propto\nu^{2}$.  In Fig.~2, the dashed line represents a power law with $\alpha=2$ normalized to the optical data. Note that the UKIRT observations (taken on 2002 February 14) are consistent with a thermal emission from the disc. 
We clearly observe a deviation of the IRAC flux density distribution from a thermal Rayleigh-Jeans spectrum. Based on the approximately $450$ sources detected in the IRAC field of view of our observations with a flux density $>$0.1 mJy, the chance probability that a mid-IR emitting source is within $4$~arcsec$^2$ of the 4U~0614+091 optical position is $\sim2\times10^{-3}$. This has to be considered a firm upper limit, given that only a sub-sample of the detected sources are likely to have the observed power-law spectral shape. The negative spectral index $\alpha$ of the power-law spectrum in the mid-IR excludes also the possibility of a circumbinary disk (e.g. Dubus, Taam \& Spruit 2002).  Therefore, the non thermal mid-IR radiation observed is synchrotron emission from relativistic electrons in a (jet) outflow.

\subsection{The break frequency}

We do not directly observe the `break frequency' $\nu_{thin}$ between the optically thick and the optically thin synchrotron spectrum, which must therefore be at frequencies lower than $3.7\times10^{13}$~Hz. This break frequency is still possibly consistent with the upper limits obtained for XTE J1118+480 (Hynes et al. 2003a), but definitely lower than that of other BH XRBs (e.g. V404 Cyg, GRS~1915+105) for which the optically thick synchrotron spectrum has been observed up to the near-IR and optical band (see Fender 2001 and references therein).  In GX~339-4, we observe a possible break frequency at a few times $10^{14}$~Hz for the lower spectrum in the right panel of Fig.~3 ({\it open circles}). This $\nu_{thin}$ is about a factor of 10 higher than the upper limit we find in 4U~0614+091. Falcke, K\"ording \& Markoff (2004) predict for compact jets in BHs that the  break frequency scales with the physical size of the jet at the base, $R_{nozzle}$, and the mass accretion rate, $\dot{M}$, as $\nu_{thin}\propto R_{nozzle}^{-1} \dot{M}^{2/3}$. If the base of the jet is linked to  the accretion disk, in BHs the {\it minimum} $R_{nozzle}$ would be the innermost stable orbit, i.e. proportional to the mass of the compact object, while in NSs the inner disk radius would depend also on the strength of the surface magnetic field. Within this framework, the observed difference in $\nu_{thin}$ between 4U~0614+091 and GX~339-4 can be accounted for by a larger $R_{nozzle}$ and/or lower $\dot{M}$ in the NS.

\subsection{Jet power}

The 2-10~keV  luminosity of the source measured from the daily-averaged count rate of the All-Sky Monitor (ASM) on-board the {\it Rossi X-Ray Timing Explorer} ({\it RXTE}) on the day of the {\it Spitzer} observation is $L_{2-10~keV}\sim1.6\times10^{36}$~erg/s (at a distance of 3~kpc). Using the X-ray bolometric correction in Migliari \& Fender (2006), i.e. for an atoll NS in hard state $L_{2-10~keV}\sim0.4 L_{Xbol}$, we obtain $L_{Xbol}\sim4\times10^{36}$~erg/s. 

Since we have an upper limit on the break frequency in 4U~0614+091, we can estimate a conservative lower limit on the total power content in the optically thick plus the observed IR part of the jet with respect to the accretion X-ray power. Given the optically thin spectrum with $\alpha=-0.57$ between 3.5 and 8 $\mu$m, and assuming (1) a flat spectrum from the radio band up to $8~\mu$m,  (2) a radiative efficiency for the jet of 5\% (Blandford \& K\"onigl 1979; see also Fender 2001, 2006),  and (3) no correction for relativistic bulk motion (Fender 2001; Gallo et al. 2003; see also Heinz \& Merloni 2004), we estimate a lower limit on the compact jet total power up to $3.5~\mu$m of $L_{j}\ge3.9\times10^{33}$~erg/s. Only the flat part of the spectrum would be $L_{j}\sim2\times10^{33}$~erg/s.  Therefore, the lower limit of the `observed' total jet power (up to $3.5~\mu$m) in this NS system is $\sim0.1\%$ of the X-ray bolometric power. Considering only the optically thick part of the spectrum, we find the total jet power lower limit is 0.05\% of the X-ray bolometric power, about 2 orders of magnitude less than the lower limits of $\sim10\%$ inferred for the optically thick spectrum of BHs with the same assumptions (Fender et al. 2000; Fender 2001; Corbel \& Fender 2002). 
Given that the optically thin part of the synchrotron spectrum in 4U~0614+091 is likely to extend to higher energies, up to a `high-energy cutoff' frequency above which the flux drops due to cooling processes (e.g. Heinz 2004;  Kaiser 2006), the slope of the optically thin emission that we have measured allows an estimate of the total power in the jet.   In Fig.~3, we show the broadband spectrum of 4U~0614+091 compared to that of the BH  GX~339-4.  Assuming an optically thin spectrum with $\alpha=-0.57$, we find  the total jet power is  $L_{j}\sim2\times10^{35}\times (E_{1~keV})^{0.43}$ erg/s (where $E_{1~keV}$ is the high energy limit of the integration in units of 1~keV). If we extend (arbitrarily) the jet spectrum up to 1~keV, the total jet power would be $\sim5\%$ of the X-ray bolometric luminosity. 
%A lower limit on the total jet power of 20\%  the X-ray luminosity, has been estimated for the BH XTE~J1118+408 integrating only up to the IR-optical band (Fender et al. 2001). Note that most of the jet power comes from the optically thin portion of the spectrum, at frequencies higher than the optical band.

While the 2-10~keV X-ray fluxes in 4U~0614+091 and in the BH GX~339-4 are comparable, the observed flux of the synchrotron component in the BH is much larger (of at least two orders of magnitude) than that in the NS system (see Fig.~3). This observed difference in $L_{radio}/L_{Xbol}$ can be explained by a difference in the role of the jet as a power output channel of the two systems: in hard-state BHs the jet could carry a dominant fraction of the accreting power (in the definition by Fender, Gallo \& Jonker 2003, with the hypothesis of no-advection, the system is `jet-dominated' when the total jet power dominates over the X-ray bolometric luminosity), while NSs never enter such a `jet-dominated' regime (Migliari \& Fender 2006).
The fact that the total jet power is a large fraction of (and possibly exceeds) the X-ray luminosity in hard state BHs still allows for the possibility of a significant advective accretion flow (see advection-dominated accretion flow : e.g. Narayan \& Yi 1994; adiabatic inflow-outflow solution: Blandford \& Begelman 1999). As discussed by K\"ording, Fender \& Migliari (2006), if the fraction of accretion power that goes into the jet is roughly the same in BHs and NSs (and the disk winds are similar for the two systems at a given Eddington-scaled accretion rate), then some of the relative dimness of BHs in the X-rays can be accounted for by significant advection of energy through an event horizon.

\section{Conclusion}

We have observed the  NS XRB 4U~0614+091 with {\it Spitzer} IRAC and obtained the following results: 

{\it i)} We detected the IR counterpart of 4U~0614+091.
 The IR spectrum is non-thermal and well fit by a power law with spectral index $\alpha=-0.57\pm0.04$. The IR radiation is the optically thin synchrotron emission from relativistic electrons in a (jet) outflow: the `break' part a compact jet spectrum. 

{\it ii)} We place a firm upper limit of $\nu_{thin}=3.7\times10^{13}$~Hz on the break frequency $\nu_{thin}$ of a NS system. This is at least a factor of 10 lower than the $\nu_{thin}$ in BHs.    

{\it iii)} For the optically thick jet, we estimate the total jet power to be $>0.05\%$ of the X-ray bolometric luminosity for 4U~0614+091, i.e. 2 orders of magnitude lower than the lower limit of 10\% inferred in BHs, using the same assumptions.  By extrapolating the optically thin jet of 4U~0614+091 up to 1 keV, we infer a total jet power corresponding to only 5\% of the X-ray bolometric luminosity.

{\it iv)} The difference in the $L_{radio}/L_{Xbol}$ ratio we find between the NS 4U~0614+091 and BHs can be explained by a difference in the role of the jet as a power output channel in the two systems: BHs are `jet-dominated', while NSs never enter a `jet-dominated' regime. There is still the possibility that in BHs, while the total jet power dominates the X-ray bolometric luminosity, a significant fraction of the total accreting power is advected through the event horizon. 

%strongly support the idea that BHs in hard state are `jet-dominated' (in the definition of Fender, Gallo \& Jonker 2003), while NSs never enter such a jet-dominated state and the (non-jet) X-ray emission is still the dominant power output channel.

\acknowledgments

The authors would like to thank Elmar K\"ording for very useful discussions and Stephane Corbel for providing us with the data of GX~339-4. S.M. acknowledges helpful discussions with Mari Polletta. E.G. is supported by NASA through Chandra Postdoctoral Fellowship Award PF5-60037, issued by the Chandra X-Ray Observatory Center, which is operated by the Smithsonian Astrophysical Observatory for NASA under contract NAS8-03060. This work is based on observations made with the {\it Spitzer Space Telescope}, which is operated by the Jet Propulsion Laboratory, California Institute of Technology, under a contract with NASA. Support for this work was provided by NASA through an award issued by JPL/Caltech.

%####################################################
\begin{figure}
\begin{center}
\includegraphics[angle=0,scale=0.8]{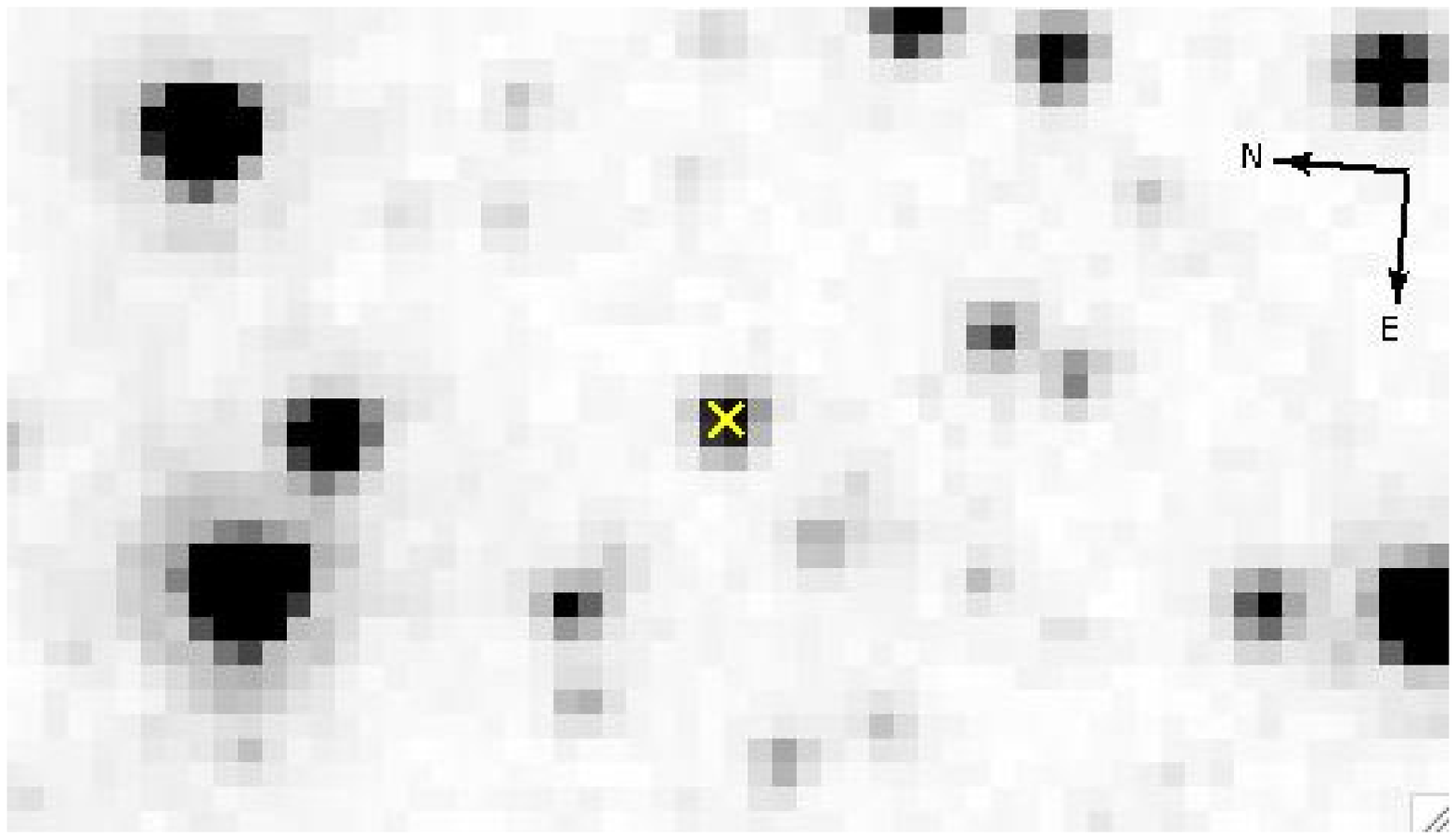}
\caption{{\it Spitzer} IRAC map at $4.5 \mu$m, centered on 4U0614+091. One pixel size corresponds to 1.2 arcsec. The cross indicates the position of its optical counterpart, V1055 Ori.}
\label{default}
\end{center}
\end{figure}

%##################################################

%##################################################
\begin{figure}
\begin{center}
\includegraphics[angle=-90,scale=0.7]{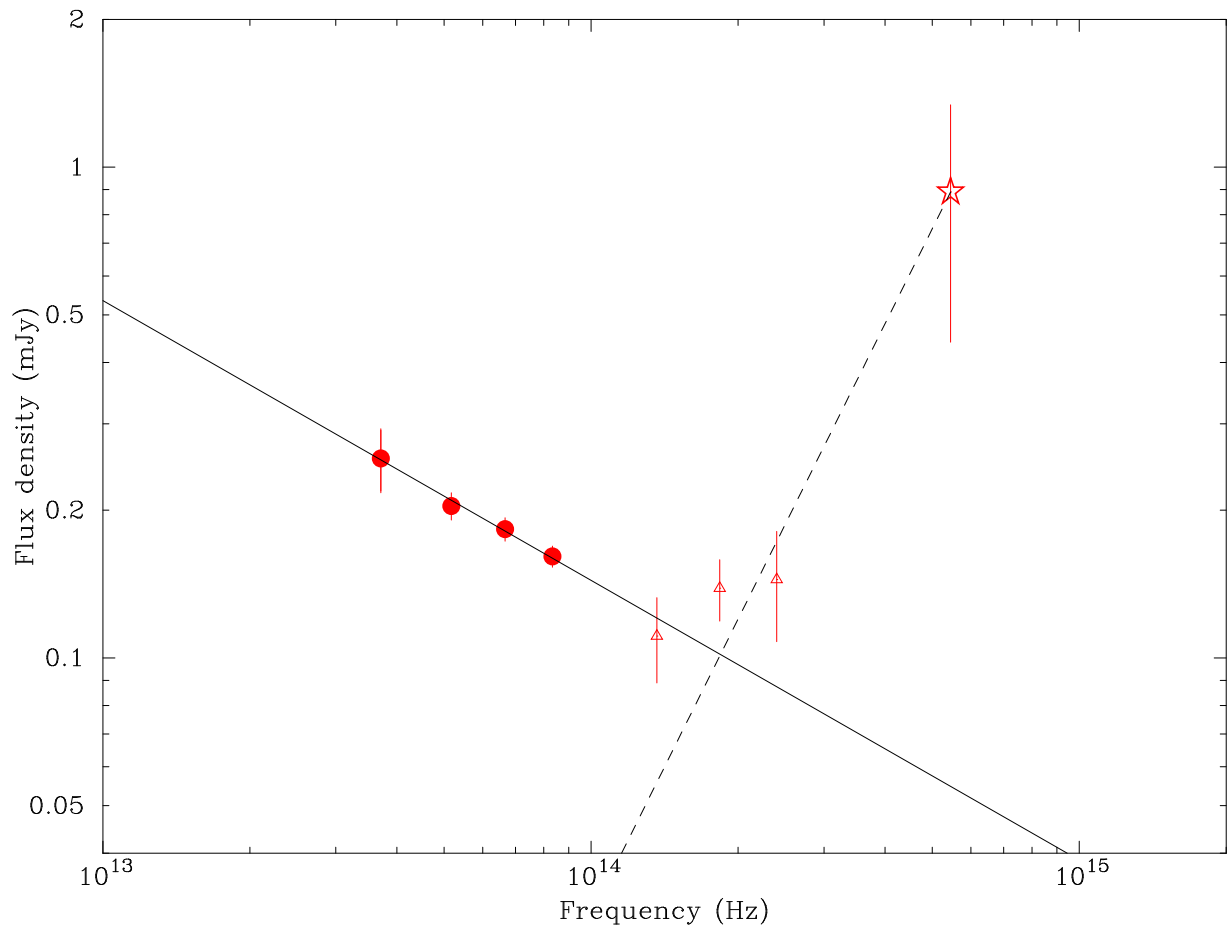}
\caption{{\it Spitzer} IRAC (filled circles), UKIRT {\it J, H, K} (open triangles) and  mean optical (star) observations of 4U~0614+091. The solid line is the fit with a power law of the IRAC data, resulting in a best-fit spectral index of $\alpha=-0.57\pm0.04$. The dashed line represents the Rayleigh-Jeans law for the thermal emission of the disc, with $\alpha=2$, normalised to the optical data. The error bar of the optical data represents the range of  the observed optical flux variations.}
\end{center}
\end{figure}
%##################################################

%###################################################
\begin{figure}
\includegraphics[angle=0,scale=0.18]{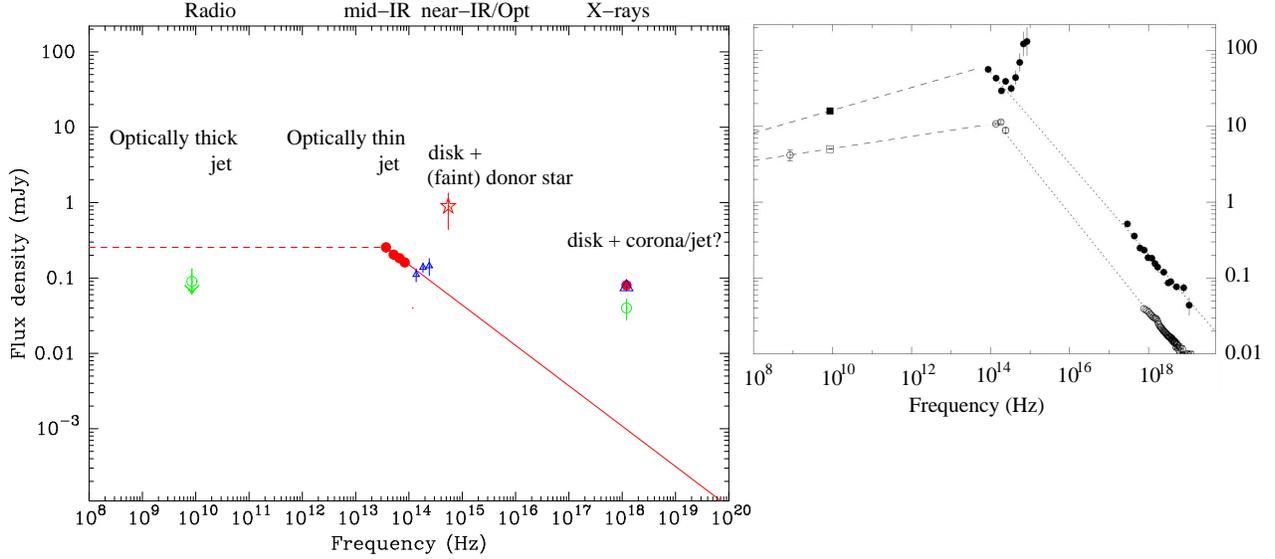}
\caption{{\it Left:} broadband spectrum of the neutron star 4U~0614+091. The filled circles are the simultaneous {\it Spitzer} IR and ASM/RXTE X-ray observations on 2005  October 25, the open circles are the simultaneous WSRT radio and ASM/RXTE X-ray observations on 2001 April 24 (Migliari \& Fender 2006) and the open triangles are the UKIRT {\it J, H, K} and ASM/RXTE X-ray observations on 2002 February 14 (Russell et al., in prep.). Note that the X-ray flux during the UKIRT observations is similar to that during the {\it Spitzer} observations. The solid line is the fit to the {\it Spitzer} data. The dashed line represents a flat optically thick spectrum normalized to the highest IR flux density. Note that, although the radio upper limit seems to indicate an inverted synchrotron optically thick spectrum, the WSRT observation is not simultaneous with the {\it Spitzer} observations and it corresponds to a lower X-ray luminosity. {\it Right:} broadband spectrum of the black hole GX~339-4 (from Corbel \& Fender 2002). Filled symbols are from observations in 2001 and open symbols are from observation in 1997. Circles are actual quasi-simultaneous observations, squares are estimated fluxes based on the radio/X-ray flux correlation found for GX~339-4 (Corbel et al. 2003; see details in Corbel \& Fender 2002).}
\end{figure}
%#############################################

\end{document}